# Towards considering digital transformation of human interactions through the lens of health

Perspective

N Bidargaddi[1], J.C.L. Looi[2,3]

**Corresponding author:** A/Prof Niranjan Bidargaddi, Digital Health Lab, College of Medicine and Public Health, Flinders University, Tonsley, GPO Box 2100, Adelaide, SA 5001, Australia.
Email: Niranjan.bidargaddi@flinders.edu.au.
Ph: (08) 7221 8840
ORCID iD: 0000-0003-2868-9260

J.C.L. Looi
ORCID iD: https://orcid.org/0000-0003-3351-6911

**Full name, department, institution, city and country of all co authors**
1. Flinders Digital Health Centre and Flinders Health and Medical Research Institute, College of Medicine and Public Health, Flinders University, Adelaide, Australia
2. Academic Unit of Psychiatry and Addiction Medicine, The Australian National University Medical School, Canberra Hospital, Canberra, ACT, Australia
3. Consortium of Australian-Academic Psychiatrists for Independent Policy and Research Analysis (CAPIPRA), Canberra, ACT, Australia.

**Abstract**

The profound intertwining of digital interactions with traditional human interactions has spawned a novel mode of human interactions at both individual and population levels, carrying substantial ramifications for health. Digital mediation refers to the process by which digital interactions influence and facilitate human interactions. While digital mediations can bring benefits, they also pose potential health risks. We identify four levels where problems may arise: the medium itself, mediation architects, end users, and mediation orchestrators. Addressing the challenges associated with shaping digital mediation of human interactions for health requires strategies and future research. Shifting focus towards understanding the impact of digitally mediated interactions on health is crucial for advancing research, policy, and practice in this field.

Word count: 2,909

Figures: 2

References: 45

**INTRODUCTION**

Although pioneered in business applications, integration of assistive digital technologies in human interactions such as work, leisure or socialisation may also influence health [1]. Interactions with health services such as medical appointments, referrals, prescriptions, health advice, etc., are also increasingly reliant on digital processes, especially since the start of the COVID-19 pandemic [2], [3], and seem likely to stay so beyond it. Human health is thus shaped by interactions, as well as the influences of digital technologies mediating them.

Contemporary health services research on health and technology [4], [5], places a greater emphasis on digital technologies as instrumental health care delivery tools, i.e., to problem-solve: diagnose, predict, detect or find information, and more recently as prevention and health promotion tools in the realms of public health [6], [7]. As digital technologies expand in societal usage outside the health services [1], [8], it is imperative to consider health implications of unfolding digital transformations of human interactions [9], [10]. These digitally-mediated interactions need to be considered through the viewpoints of ethical [11], autonomy-preserving [12] and equitable-access[13], [14] design principles. This is an important issue as digital applications are not purely instrumental, but are experienced as an extension of self-agency [15]–[17].

Previously, the implications of technology-mediated communication on human health has been explored previously in a limited manner, constrained by narrower conceptualisations of usage [18]–[20]. This prior research has focused on understanding healthcare effects by considering in isolation either the technology or the human as the responsible actor, and less so on conceptualising the fully integrated interaction of technology and humans in the mediation of health outcomes. To address this gap, we propose that, as technologies permeate various aspects of human life, an integrated multi-component digitally mediated human interaction becomes a potential mechanism for enhancing individual and collective health. We then discuss the challenges of shaping health in digitally transformed human interactions.

**UNDERSTANDING DIGITALLY MEDIATED HUMAN INTERACTION FROM HEALTH PERSPECTIVE**

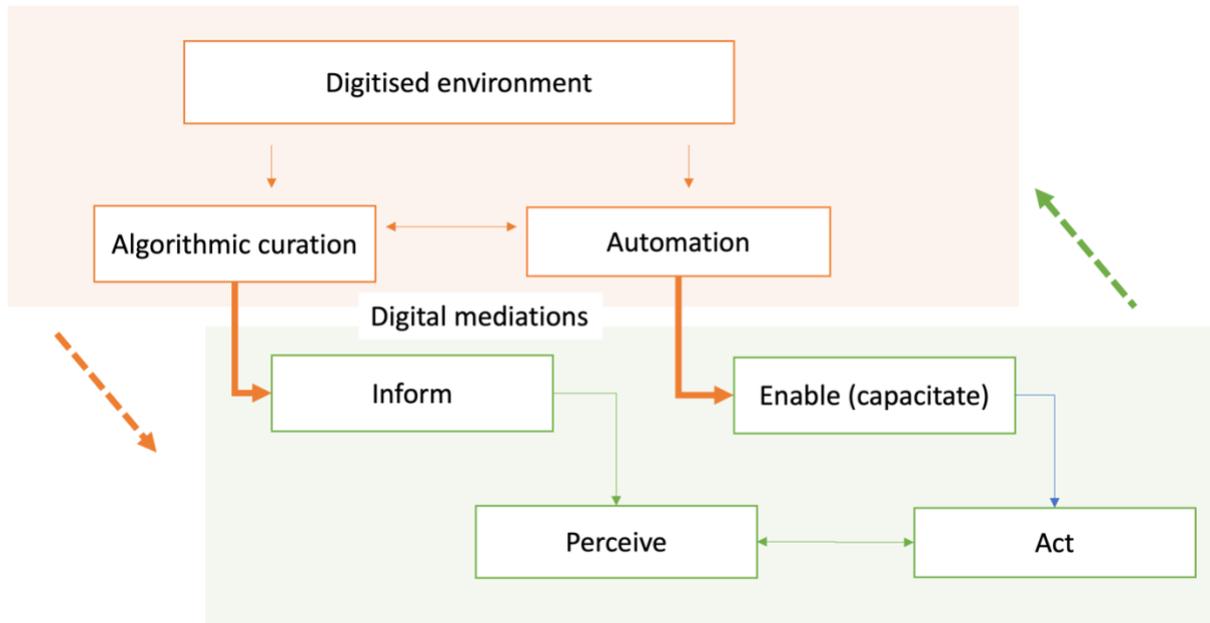

**Figure 1 –Digital mediation process**

Digital mediation is the process of interweaving technologies into human daily life to tackle constraints on time and cognitive resources. An overview of social theories explored within studies of technologies mediating interactions within health care [21], [22] suggests when technologies are introduced to mediate a human interaction, new frames for comprehending and responding to situations in time and space are created, which imply new ways of organising and making sense of experience, and require effort by the participants in the interaction[23], [24]. Others have argued that the experiences in a mediated interaction is shaped both by normative framing expectation of interaction and the affordance offered by the new medium [25], [26].

From a mechanistic perspective, human decision-making comprises two major cognitive systems, a fast, heuristic-based system, and a slower, more analytical process [27]–[29]. As humans are innately evolved and culturally-shaped to function as part of cooperative framework [30]–[32], interactions by people (decisions, actions) are not static, but instead are dynamic, relational, and contextual responses at individual and social levels. While human perceptions are mainly shaped by information one receives, it also depends on capability (such as time, energy, skills, resources [33]–[35]) when action is expected to occur. Such complex cognitive and cooperative processes behind human decisions, actions and unfortunately omissions, as well as accessibility, affordability and effectiveness of interventions and resources, equally determine health and wellbeing. As digital mediations replace or augment whole or part of what would have been otherwise a human task or decision, these complex health shaping processes face interference.

From a digital transformation perspective, providing timely information and enhancing capability are the reasons for digitally mediating a human interaction. Different types of digital technologies are being integrated in different contexts at individual and population level to enhance these needs for information and capability across the lifespan (Figure 1). Based on functionality (Figure 1), the two common categories of digital technologies are algorithms (which provide information that shapes perception) and automations (which substitute for what would have been otherwise labour or action exerted by humans)[25]–[28]. It is important to consider these technologies as mediators, as they don't just transport meaning or force without transformation, but modify the meaning of the elements they are supposed to carry[36], [37].

Drawing on these perspectives, we conceptualise a digitally mediated human interaction as comprising four components: (i) Digital technology (interface over which the mediating transaction is actualised), (ii) Designer (a human configuring the mediation), (iii) Receiver (a human at the experiencing end of interaction) and (iv) Provider (a human at the instigating end of the interaction). When a digitally mediated human interaction adversely impacts health, the reason or cause for that may arise from constraints, uncertainties, and errors within one or more of these components. These issues are examined from the perspective of each stakeholder or component, next.

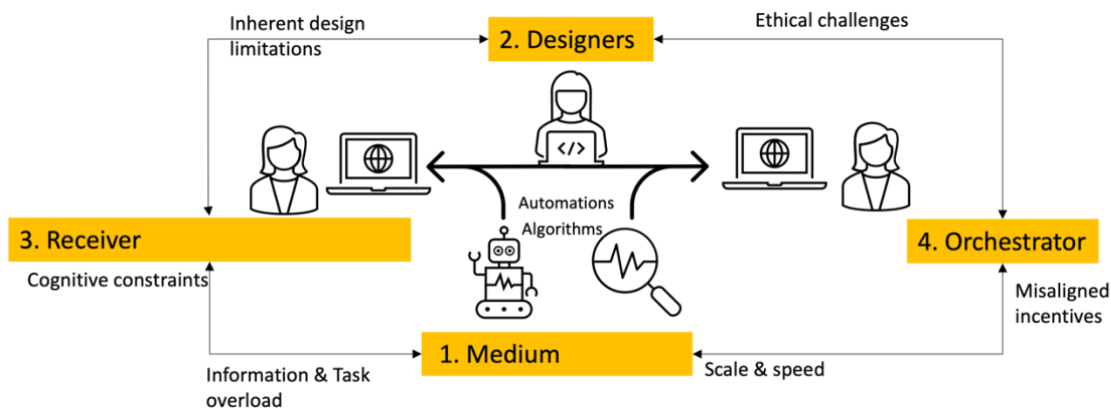

**Figure 2: Composite parts of a digitally mediated human interaction**

**THINGS THAT GO WRONG WITH SHAPING A DIGITALLY MEDIATED HUMAN INTERACTION**

**(i)      Hazards of mediating technology**

Digital mediations are not simply instrumental tools as they operate primarily in the realm of cognition, i.e., they are not just interactive mechanisms. Such technology can present benefits as well as challenges.  Firstly, digital mediation can only partially mimic humans because a mechanical digital interface however advanced, is in effect designed to meet specified requirements, an approximation of explicitly specifiable observations and experiences that underpin an interaction.  Digital interfaces are not well equipped to deal with uncertainty, in contrast with humans. Human interactions are often comprised of a priori unspecifiable emergent responses predicated on many factors such as cognition, emotion and circumstance. For example, a conversation between a doctor and patient unfolds not just in speech, but also in subtle signs of body language, facial expression, tone, etc that allows each party to adjust, and thus alter the flow of communication in a stochastic way[24], [38].  To date, automated tools have not been able to conduct a realistic human conversation beyond some artificially constrained scenarios [39].

Secondly, digital mediations mimicking human functions (anthropomorphism) are not equipped to comprehend and process the nuances of uncertainty in ways human do. For example, consider a virtual emergency triaging gateway interface into which people enter their symptoms and receive a response on actions they should take next, as an alternative to people presenting to Emergency Department. Suppose a person reports chest pain - the algorithm in the gateway, constrained to its programming, may ask follow up questions to establish severity, whereas another critical question is the nature of the chest pain as well. Depending on this response the algorithm may respond back either a less urgent or a critical action as the next step. However, one type of pain may have represented indigestion, and the patient may have minimised the severity of a pain that resembled a cardiac cause. Were these interactions happening in person this necessary contextual detail is less likely to go unnoticed. When people make mistakes liaising with the digital mediation, irrespective of the reason, which may be intentional or for accidental reasons, in a health context consequences can be drastic [40], [41].

Third, as digital mediations can scale to populations quickly, they can also inadvertently scale incorrect or ineffective interactions efficiently. Interacting with a digital mediation is may be analogous to collaborating remotely with another person to achieve a common goal which may involve synchronous or asynchronous processes. However, the digital mediation only partly resembles human interactions. In these kind of virtual team interactions errors can arise for different reasons [42]. For example, temporal misalignments occur when there is an error due to delays or slow response at receiving or sending side of an interaction. The following illustrates an example of this problem. A health professional using emails to communicate with patients may organise workflow such that they review messages from patients at a fixed time once every week. This could equally be an AI tool. A patient may email critical information immediately after the review time, and as a result may not receive the next response until a week later when their situation may have warranted an immediate response. As the order, sequence, depth of information flow and asynchronism increases within an interaction mediated by automation or algorithms, errors may further accumulate at each step further diminishing the value of mediation.

**(ii) Designers' values and biases**

Digital mediation architectures are shaped by designers according to their understanding of requirements. Software developers (programmers, designers, etc) receive requirements that do not capture all that is necessary to define a human interaction because of the very inherent limitations of requirement gathering process[43], [44]. There are a wide variety of implicit, nuanced and unforeseen situations in human interaction which cannot be observed and specified reliably. The separation between software developers (programmers, designers) and the people (users) they are designing for across time, space and other contextual axis acts as a barrier to understanding the true intentions of their creation. Even assuming software developers (programmers, designer, etc) completely understood the nuances of mediated interactions, misconfiguration can occur because of other reasons attributable to values and identity [45]–[47]. For instance how Health Professionals make decisions are underpinned by solid ethical principles [11], while professional standards and regulatory oversight enforces their adherence. Ethical frameworks for the software industry as a whole are not enforced [48], instead individual software professionals bring their own values and biases into the design process, and these may significantly differ from those of health professionals, as well as healthcare consumers, using the interfaces. The definitions of inherent values are not necessarily the same [49], [50].  Equally important is ownership – who controls and pays.  While configurable options are determined by developers, default configurations are usually set to suit the preferences of administrators or distributors[51].

Given the prominent role of software professionals in shaping society distally through digital-mediation, ethical training, professional and regulatory oversight commensurate to those for health professionals could be enacted. Incorporating healthcare ethics into guiding principles for software development will facilitate design of digital-mediations that are ethically fit-for-purpose.

### (iii)    People experiencing digital mediation

For whatever reason, if data received through an algorithm is not sufficiently informative, or when an automation is not useful; from the perspective of a person at the receiving end, such mediations are ineffective noise, and may erode capacity to comprehend or act further [52]–[54]. For example, health professionals now face potential information overload from Electronic Health Records offering granular data on patient history and algorithms providing multiple sources of data, diagnoses, as well as risk profiles, and also, sometimes suggested actions [52]–[54]. Concurrently, health professionals face enormous cognitive, time and resource constraints due to the real-world need to attend to critical time-sensitive care of patients [52]–[54]. They can pay attention to only that is critical or urgent.  Increasing volumes of patient information, however potentially rich and predictive, may entropy into noise if it cannot be parsed for comprehension and response within the brief available time [52]. Similarly, there are numerous, difficult to discern and navigate health support resources on internet of differential usefulness for consumers, and such information overload may lead people towards adverse information and choices [55]. Another example is the app market place which has over a million health apps with many of poor design quality and unknown or doubtful healthcare efficacy. Finding safe and effective options quickly is a challenge likewise. Inevitably people may need to be educated to develop skills and acquire resources to adapt to the demands and constraints of digitally-mediated interaction [56], [57].

**(iv) Orchestrators of digital mediation**

A digitally-mediated human interaction is orchestrated by or on behalf of a 'human' provider — a person who has ethical-legal responsibility for the consequences of what is offered or done, however diffused and distal they may appear. Depending on the type of mediation instigated, a provider may range from a health service agency that is relying on algorithms and automations to an individual practitioner using digital-mediation mainly to facilitate face-to-face or telehealth consultations.

Sometimes digital mediation may involve an employer using digital automations in a routinised workflow. This may also include online services people use in daily life for socialising, entertainment, shopping etc. There is the potential and reality that digital mediations may be developed against users' interests by design because of the commercial or other interests of the mediation-provider. These problems are exemplified in how everyday technologies people use are optimised to keep people engaged with digital offering – visit them more or spend more time on them, even though excessive use can harm wellbeing and productivity downstream [58]. For example, when an app is installed on a mobile phone by default the notification function is enabled to enhance attention to the app and its output [59]. Sometimes, in social media newsfeeds or search engines, users are not able to directly influence or redirect the algorithm that is serving them information [60], [61]. Efficiently streamlined manufactured consent processes direct people away from knowledge of this covert influence [51]. Programming of digital mediations may thus operationalise erosion of personal ethics and agency [51]. There are many examples of manufactured consent and surveillance processes that alienate populations [51], [62]. Finally, digitally-mediated outsourcing to an external automated agent partial thinking and acting processes risks weakening/eroding human agency and autonomy [51], [62].

**DISCUSSION**

The trend towards shaping and roll out of digitally mediated products, processes and services within business sectors (including health services) has accelerated rapidly since the COVID-19 pandemic. Digitally mediated products, processes and services in turn are now prominently shaping, either directly or indirectly, many human interactions in daily life. Digital mediations aim to either inform (i.e., algorithms) or enhance capability (i.e., automation) of human interactions such as work, socialising, entertainment, recreation, healthcare, etc. While such digital mediations may lead specific experiential enhancements (specific-effects), they may also have an indirect effect on overall health (non-specific effects), since such daily interactions in turn, directly or indirectly impact on human health.

Digital mediations (algorithms or automations) have specific effects in relation to what they were designed to achieve (i.e., shop efficiently) and non-specific effects in relation to overall health (i.e., shopping for healthy food or related products). How digital mediations are designed and set up are an important aspect of health interactions. We outlined four different levels at which problematic health interactions may occur: (i) the digital technology to facilitate the health interaction, (ii) the human programmers/designers developing the digital mediation technology, (iii) people receiving care through the mediated interaction and (iv) people providing healthcare through the digitally-mediated interaction.

**Potential solutions**

1. Embedding design consideration of the inherent anthropomorphic limitations in developing digital mediations is critical to preventing ineffective mediation of healthcare interactions.

2. Appraisal of overall aims, values, and health targets can steer developers and designers develop digital-mediation solutions that are fit-for-purpose and also beneficial to health of society.

3. Developers and designers need to incorporate professional education, ethics and practice cognate to that for healthcare professionals if they are to design digital mediations for healthcare interactions.

4. Developers and designers will need to work in co-design with healthcare consumers and providers for effective digital mediation in healthcare interactions.

5. Digital competency should become part of curriculum in school education particularly focussing on teaching people how to interact with digitally mediated products and services, and especially healthcare.

6. There are values and biases inherent to the protocols which bring technology mediations into life and manage information flow. The dominant protocols so far have favoured centralised control architectures (i.e., small number of people having greater control in defining rules for the majority of the users). Decentralised model (i.e., distributed control amongst users) based on alternative architectures such as blockchain can delineate platform ownership (i.e., finance) from control (who govern and configures), but may also introduce the risk of manipulation by users or providers that are technologically capable (e.g., targeted hacks to steal funds from cryptocurrency exchanges).

**Future directions**

We have not discussed here the role of the cognitive science of human decision-making, and the related field of choice architecture, as our focus is primarily upon the digital mediation of healthcare interactions. However, these matters warrant further consideration in related work and research, which may be considered in a separate paper. Similarly, the specific protocols for control and use of digital mediations are also relevant, and worthy of separate discussion. Finally, there remains the issue of fairness and equity in the ethical provision of healthcare which also can be affected by digital mediation.

**CONCLUSION**

Digital products, services and processes have the potential to significantly improve healthcare interactions. However, careful co-design and development with users and healthcare providers will be needed to promulgate clinically effective mediation. Comprehensive design of digital mediations will need to consider, and avoid pitfalls at a number of levels: (i) the digital technology to facilitate the health interaction, (ii) the human programmers/designers developing the digital mediation technology, (iii) people receiving care through the mediated interaction and (iv) people providing healthcare through the digitally-mediated interaction.

## Contributions

NB developed the concept for the manuscript, wrote the article, and reviewed the final content. JL contributed to the conception and editing of the manuscript.

## Competing interests

NB has shares in goAct which has a license from Flinders University for the MindTick platform and has received funding from ARC Australian Industry Transformation Hub and Digital Health CRC for collaborations with goAct. JL declares no competing Financial or non-Financial Interests